\documentclass[12pt]{iopart}

\usepackage{graphicx}
\begin{document}

\title{Pre-scission neutron multiplicity associated with
the dynamical process in superheavy mass region}

\author{Yoshihiro Aritomo$^{1}$, Masahisa Ohta$^{2}$ and Francis Hanappe$^{3}$}


\address{$^{1}$ Flerov Laboratory of Nuclear Reactions, JINR, Dubna, Russia \\
$^{2}$ Department of Physics, Konan University, 8-9-1 Okamoto, Kobe, Japan \\
$^{3}$ Universite Libre de Bruxelles, 1050 Bruxelles, Belgium}

\ead{aritomo24@muj.biglobe.ne.jp}

\begin{abstract}
The fusion-fission process accompanied by neutron emission is
studied in the superheavy-mass region on the basis of the
fluctuation-dissipation model combined with a statistical model. The
calculation of the trajectory or the shape evolution in the
deformation space of the nucleus with neutron emission is performed.
Each process (quasi-fission, fusion-fission, and deep quasi-fission
processes) has a characteristic travelling time from the point of
contact of colliding nuclei to the scission point. These dynamical
aspects of the whole process are discussed in terms of the
pre-scission neutron multiplicity, which depends on the time spent
on each process. We have presented the details of the
characteristics of our model calculation in the reactions
$^{48}$Ca+$^{208}$Pb and $^{48}$Ca+$^{244}$Pu, and shown how the
structure of the distribution of pre-scission neutron multiplicity
depends on the incident energy.

\end{abstract}

\maketitle







\section{Introduction}

Recently, spectacular news on the finding of superheavy elements has
been successively reported from FLNR, GSI and RIKEN
\cite{ogan991,ogan01,ogan04,hofm00,mori03,mori04}. In theoretical
approaches, the clarification of the reaction mechanism in the
superheavy mass region and the prediction of the favorable
conditions (projectile-target combination and beam energy values) to
synthesize new elements have advanced step by step in the last few
years \cite{zag,nas,dia,swi,ada,mis,rum}. However, we are still
seeking suitable models for calculating the fusion and the survival
probabilities, and investigating the unknown parameters involved in
them. The fusion-fission mechanism in the superheavy-mass region
must be elucidated; in particular, a more accurate estimation of
fusion probability should be established.

Generally, on the base of analysis of the experimental mass, kinetic
energy, and angular distributions of the reaction fragments, the
whole reaction process in heavy nucleus collisions is classified
into the fusion-fission, quasi-fission and deep inelastic collision
processes and so on. Generally, mass symmetric fission fragments
have been considered to originate from a compound nucleus.
Therefore, in the experiment, the fusion-fission cross section has
been derived by counting mass symmetric fission events
\cite{itki01}.

However, in our previous work \cite{ari04}, we showed that it is
ambiguous to identify mass symmetric fission events as the
fusion-fission process in the superheavy-mass region. Using the
Langevin equation, which enables to calculate the time development
of a trajectory of a colliding system in a shape parameter space,
we attempted to distinguish different dynamical paths by analyzing
the trajectories and mass distribution of fission fragments. We
identified the fusion-fission (FF) path, the quasi-fission (QF)
path and the deep inelastic collision (DIC) path. In addition, in
the reaction $^{48}$Ca+$^{244}$Pu, we found a new path, called the
deep quasi-fission (DQF) path, whose trajectory does not reach the
spherical compound nucleus but goes to the mass symmetric fission
area.
The conclusion is that the analysis of only the mass distribution
of fission fragments is insufficient to identify the reaction
process described between the FF and DQF paths. To estimate the
fusion probability accurately, the precise identification of the
FF path is very important.

On the basis of experimental results, it has also been suggested
that the reaction process may be classified by the measurement of
the pre-scission neutron multiplicity, which correlates with the
mass distribution of fission fragments \cite{itki01,dona99,mate04}.
This means that each process is expected to have its own
characteristic reaction time, which is related to the pre-scission
neutron multiplicity.

Hence, to identify the reaction process more precisely, we
undertake to extend our model discussed in reference \cite{ari04},
by taking into account the effect of neutron emission along the
dynamical path. We introduce the effect of neutron emission into
the three-dimensional Langevin calculation code, or we combine the
fluctuation and dissipation dynamical model with a statistical
model. The preliminary analysis along this line has been done in
the reaction system $^{58}$Ni+$^{208}$Pb at a incident energy
corresponding to the excitation energy of the compound nucleus
$E^{*}= 185.9$ MeV, which is a high incident energy case
\cite{ari04a1}.

In this study, we have applied our model to the investigation of the
whole dynamical process in the reactions $^{48}$Ca+$^{208}$Pb and
$^{48}$Ca+$^{244}$Pu mainly at $E^{*}= 50$ MeV, that is to say, a
low incident energy case. These reactions are chosen in order to
precisely investigate the fusion process in the hot fusion reaction
that has recently been performed at FLNR \cite{oga04a}.  Also we
discuss how the characteristic structure of the neutron multiplicity
depends on the incident energy.

In section~2, we explain our framework for the model used here. We
discuss the pre-scission neutron multiplicity with respect to
fission fragments in section~3. The mechanism of the
fusion-fission process with neutron emission is investigated. In
section 4, we present a summary and further discussion.

\section{Model}


Using the same procedure as described in reference \cite{ari04} to
investigate the whole process dynamically, we use the
fluctuation-dissipation model and employ the Langevin equation. We
adopt the three-dimensional nuclear deformation space given by
two-center parameterization \cite{maru72,sato78}. The three
collective parameters involved in the Langevin equation are as
follows: $z_{0}$ (distance between two potential centers),
$\delta$ (deformation of fragments) and $\alpha$ (mass asymmetry
of the colliding nuclei); $\alpha=(A_{1}-A_{2})/(A_{1}+A_{2})$,
where $A_{1}$ and $A_{2}$ denote the mass numbers of the target
and the projectile, respectively. We assume that each fragment has
the same deformations as the first approximation. In the present
calculation, the neck parameter $\epsilon$ is fixed to be 1.0, so
as to retain the contact-like configuration more realistically for
two-nucleus collision. The definitions of the $\delta$ and
$\epsilon$ are mentioned in reference \cite{maru72,sato78,ari04a}.

The multidimensional Langevin equation is given as
\begin{eqnarray}
\frac{dq_{i}}{dt}&=&\left(m^{-1}\right)_{ij}p_{j},\nonumber\\
\frac{dp_{i}}{dt}&=&-\frac{\partial V}{dq_{i}}
                 -\frac{1}{2}\frac{\partial}{\partial q_{i}}
                   \left(m^{-1}\right)_{jk}p_{j}p_{k}
                  -\gamma_{ij}\left(m^{-1}\right)_{jk}p_{k}
                  +g_{ij}R_{j}(t),
\end{eqnarray}
where a summation over repeated indices is assumed. $q_{i}$
denotes the deformation coordinate specified by $z_{0}$, $\delta$
and $\alpha$. $p_{i}$ is the conjugate momentum of $q_{i}$. $V$ is
the potential energy, and $m_{ij}$ and $\gamma_{ij}$ are the
shape-dependent collective inertia parameter and dissipation
tensor, respectively. A hydrodynamical inertia tensor is adopted
in the Werner-Wheeler approximation for the velocity field, and
the wall-and-window one-body dissipation is adopted for the
dissipation tensor \cite{bloc78,nix84,feld87}. The detail is
explained in reference \cite{ari04}.

The intrinsic energy of the composite system $E_{int}$ is
calculated for each trajectory as

\begin{equation}
E_{int}=E^{*}-\frac{1}{2}\left(m^{-1}\right)_{ij}p_{i}p_{j}-V(q,l,T),
\end{equation}

where $E^{*}$ denotes the excitation energy of the compound
nucleus, and is given by $E^{*}=E_{cm}-Q$ with $Q$ and $E_{cm}$
denoting the $Q-$value of the reaction and the incident energy in
the center-of-mass frame, respectively.


We take into account neutron emission in the Langevin calculation
during the reaction process. The emission of neutrons which is
calculated by the statistical model has been coupled to the
three-dimensional Langevin equation. The proceeder is explained
precisely in reference \cite{ari04a1}.

\section{Reaction process and pre-scission neutron multiplicity}

Our model is applied to the reaction systems $^{48}$Ca+$^{208}$Pb
and $^{48}$Ca+$^{244}$Pu with the incident energy corresponding to
the excitation energy of the compound nucleus, $E^{*}$=50 MeV. In
our previous work \cite{ari04}, we have shown that our model
calculations for the mass distribution of fission fragments agree
well with the experimental data and also the fusion-fission cross
section. On the basis of the results given in reference
\cite{ari04}, we want to make a new trial to clarify the
relationship between the dynamical process and the pre-scission
neutron multiplicity, and to investigate whether the relation can
be seen at $E^{*}$=50 MeV. These reactions are chosen bearing in
mind the more precise estimation of the fusion probability in the
so-called hot fusion reaction performed at FLNR in experiments on
superheavy element synthesis \cite{oga04a}.


In the preliminary investigation in reference \cite{ari04a1}, for
the purpose of checking the applicability of our model, we have
analyzed to the recent experiment, in which the pre-scission
neutron multiplicity correlated with the mass distribution of
fission fragments has been measured in the reaction
$^{58}$Ni+$^{208}$Pb at the incident energy corresponding to the
excitation energy of compound nucleus $E^{*}= 185.9$ MeV. This
experiment was done by D\'{e}MoN group \cite{dona99}. Our model
calculation has reproduced the characteristic trend of the
experimental distribution of pre-scission neutron multiplicity,
and distinguished the FF and QF processes among the mass symmetric
fission events.

Also we have showed our preliminary results on this study in
references \cite{mate04,ari04a2,ari04a3,ari04a0}. We investigated
the neutron multiplicity in the reaction $^{48}$Ca+$^{244}$Pu at
$E^{*}$=40 MeV. The experiment of neutron multiplicity shows the two
peaks near 2 and 4 neutron emission \cite{mate04}. Such two peaks
are originated by two different fusion-fission mechanism. It looks
that the first peak comes from QF process and the second one comes
from mainly DQF or FF. However, the calculation could not reproduce
the two peaks \cite{ari04a0}. Due to a low excitation energy, the
lifetime of neutron emission is rather short. It is comparable to
the time scale of the fluctuation of the trajectory.
So we can not distinguish QF and DQF process
clearly. Moreover the number of events of FF process is quit small
at a low incident energy. The statistics of each event are not
enough to analyze the process.


Here, we investigate further by applying in the reaction
$^{48}$Ca+$^{208}$Pb and $^{48}$Ca+$^{244}$Pu, more precisely, with
increasing the excitation energy $E^{*}=50$ MeV. To discuss the
dynamical process, we mainly focus on zero angular momentum case.
For each angular momentum, the potential landscape changes but the
mechanism of the process essentially does not change \cite{ari04a}.


\subsection{Reaction $^{48}$Ca+$^{208}$Pb}

Using the model presented in Sect.2, we calculate the pre-scission
neutron multiplicity emitted from the reaction system
$^{48}$Ca+$^{208}$Pb at $E^{*}= 50 $ MeV for angular momentum $l=0$.
In this system, the FF process is dominant, that is to say, the mass
symmetric fission events are dominant.

\begin{figure}[h]
\centerline{
  \includegraphics[height=0.39\textheight]{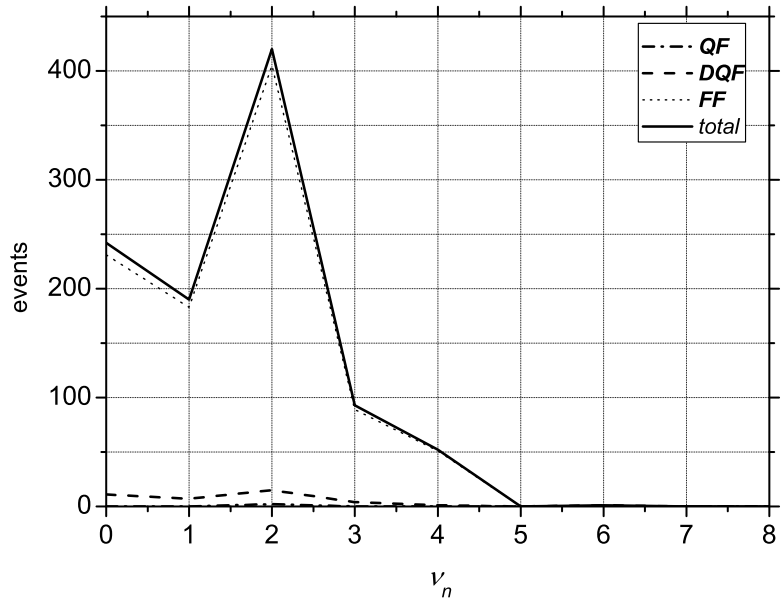}}
  \caption{Pre-scission neutron multiplicities given by our theoretical calculation
   in the reaction $^{48}$Ca+$^{208}$Pb at $E^{*}$=50 MeV for $l=0$. The
   neutron multiplicity from the QF, DQF and FF processes are
   denoted by the dashed-dotted, the dashed and the dotted lines, respectively.
   The solid line
   denotes the total process.}
\end{figure}


The calculation is done in the same procedure as reference
\cite{ari04a,ari04a1}. In the three-dimensional Langevin
calculation, we prepare 1,000 trajectories for each case. At
$t=0$, each trajectory starts from the point of contact. We define
the fusion box as the region inside the fission saddle point, $\{z
< 0.8, \delta < 0.3, |\alpha | < 0.3\}$ \cite{ari04}. Samples of
the trajectory of the FF process projected onto $z-\alpha$
$(\delta=0)$ plane in the reaction $^{48}$Ca+$^{208}$Pb at
$E^{*}=50$ MeV for $l=0$ is shown in Fig.~1 in reference
\cite{ari04a}.
The FF trajectory makes up 96\% of all trajectories, and 95\% of
the FF trajectory escapes from the spherical region within $5.0 \times
10^{-19} $ sec.

Figure~1 shows the distribution of the pre-scission neutron
multiplicity $\nu_{n}$ in this reaction. The multiplicities from
the QF, DQF and FF processes are denoted by the dashed-dotted
line, dashed line and dotted line, respectively. The solid line
shows the total multiplicity of each process. The distribution of
pre-scission neutron multiplicity has a peak around
$\nu_{n}=2$. Since the FF process is dominant, the peak
corresponds to the FF process.


We define the travelling time $t_{trav}$ as a time duration during
which the trajectory moves from the point of contact to the
scission point. We present the distribution of the travelling
times for the QF, DQF and FF processes in Fig.~2, denoted by the
dashed-dotted line, dashed line and dotted line, respectively. The
solid line shows the  distribution of the travelling time for all
processes. In the FF process, as discussed in reference
\cite{ari04a}, the trajectory is trapped in the pocket around the
spherical region. Consequently, the trajectory spends a relatively
long time in the pocket before it finally escapes from there. The
travelling time of the FF process has a peak at $1.2 \times
10^{-20}$ sec, and the average time of it is $15.3 \times
10^{-20}$ sec.

\begin{figure}
\centerline{
  \includegraphics[height=.37\textheight]{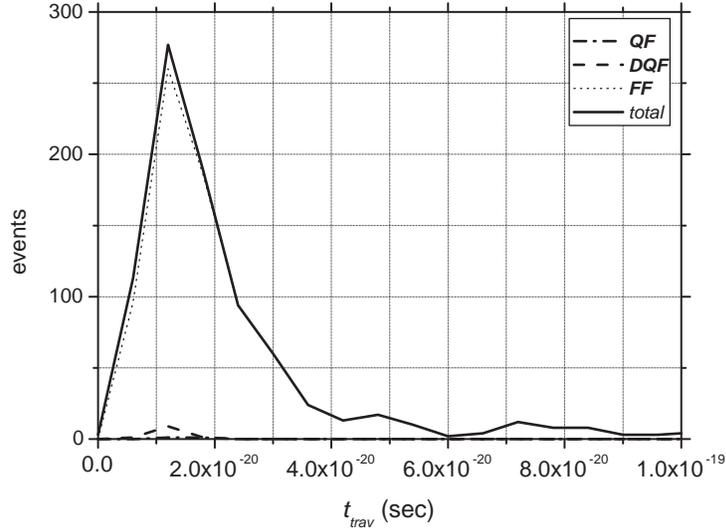}}
  \caption{The distribution of travelling time $t_{trav}$ in the
   reaction $^{48}$Ca+$^{208}$Pb at $E^{*}=50$ MeV for $l=0$.
   The $t_{trav}$ from the QF, DQF and FF processes are
   denoted by the dashed-dotted line, dashed line and dotted line, respectively.
   The solid line denotes the total process.
   }
\end{figure}

As shown in reference \cite{ari04a1}, in the reaction system
$^{58}$Ni+$^{208}$Pb at $E^{*}= 185.9$ MeV, the distribution of
the neutron multiplicity has two peaks which come from the QF and
FF processes, respectively. On the other hand, Fig.~1 shows the
single peak.

For the $^{48}$Ca+$^{208}$Pb reaction, the FF process is dominant in
this energy region, in consistent with the potential energy surface
shown in Fig.~1 in reference \cite{ari04a}, in which we can see the
steep valley in the entrance mass asymmetry and in the mass
symmetric fission valley. This fact is consistent with the
pre-scission neutron multiplicity having only a single peak at
$\nu_{n}$=2, which indicates that only one process is dominant.

In the preliminary calculation \cite{ari04a0}, we have compared the
results with the experimental data at $E^{*}=40$ MeV \cite{mate04},
which shows the distribution of the pre-scission neutron
multiplicity in correlation with fission fragments whose mass number
is grater than $\frac{A_{CN}}{2}-30$ and less than
$\frac{A_{CN}}{2}+30$, where $A_{CN}$ denotes the mass number of the
compound nucleus. The tendency of these results coincides with the
experimental data.

\subsection{Reaction $^{48}$Ca+$^{244}$Pu}

In reference \cite{ari04,ari04a}, the whole dynamical process in
the reaction $^{48}$Ca+$^{244}$Pu was classified into the QF, DQF
and FF processes. At a low incident energy, the mass asymmetric
fission events are dominant, which is the QF process
\cite{itki01}. We have distinguished between the DQF and the FF
process by analyzing whether the trajectory enters the fusion box
or not. The analysis only of the mass distribution for fission
fragments is insufficient to distinguish between the DQF and the
FF processes \cite{ari04}. In the present study, we show how each
process corresponds to a characteristic number of pre-scission
neutron multiplicities at $E^{*}=50$ MeV.

Samples of the trajectory of the QF, DQF and FF process projected
onto $z-\alpha$ $(\delta=0)$ and $z-\delta$ $(\alpha=0)$ plane at
$E^{*}=50$ MeV for $l=0$ in the reaction $^{48}$Ca+$^{242}$Pu are
shown in Fig.~5 in reference \cite{ari04a}. We define the fusion
box as the region inside the fission saddle point, $\{z < 0.6,
\delta < 0.2, |\alpha | < 0.25\}$ \cite{ari04}.

\subsubsection{Classification on the basis of mass distribution of the fission fragments}

Generally, it is considered that the reaction process is classified
according to the mass distribution of fission fragments and their
total kinetic energy distribution \cite{itki01}. The QF process is
defined as the fission fragments having mass number greater than
$\frac{A_{CN}}{2}+20$ and less than $\frac{A_{CN}}{2}-20$.
This condition is also suited to the deep inelastic collision
process (DIC) in the present case. The FF process is defined by the
condition that the trajectory enters the fusion box. The DQF process
produces the mass symmetric fission fragments (greater than
$\frac{A_{CN}}{2}-20$ and less than $\frac{A_{CN}}{2}+20$), but the
system does not reach the spherical compound nucleus before
undergoing fission. We name this method of classification as the
"classification by fission fragments mass" (standard
classification).

As discussed in reference \cite{ari04}, we can not distinguish
between the FF and the DQF processes only using the fission
fragment mass. The difference of the both processes is the
reaction time, that is to say, the travelling time $t_{trav}$. We
suppose that the neutron multiplicity depends on the $t_{trav}$.
Here, we try to investigate the correlation between the neutron
multiplicity and each process which is separated by using the
standard classification.

\begin{figure}[h]
\centerline{
  \includegraphics[height=0.39\textheight]{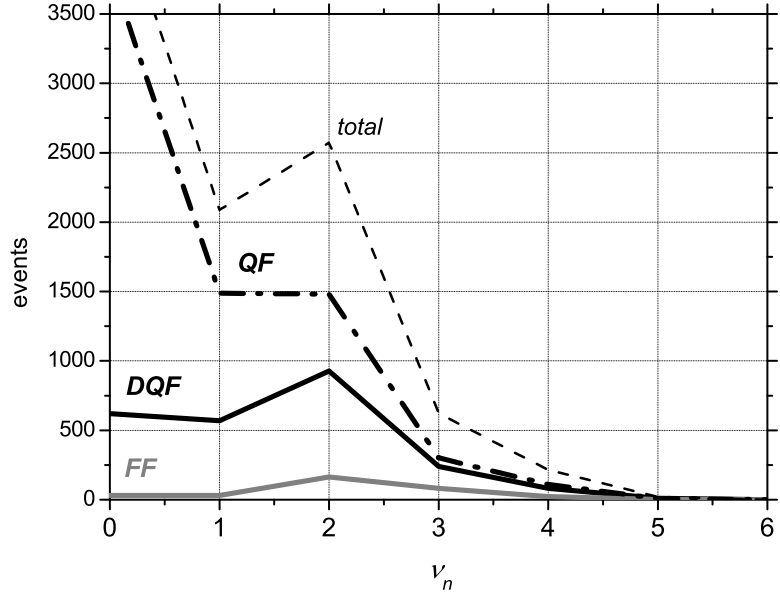}}
  \caption{Pre-scission neutron multiplicity given by our theoretical calculation
   in the reaction $^{48}$Ca+$^{244}$Pu at $E^{*}$=50 MeV for $l=0$ with the classification method
   of using the mass distribution of fission fragments. The
   neutron multiplicity from the QF, DQF and FF processes are
   denoted the dashed-dotted, the solid and the gray lines, respectively.
   The dashed line denotes the total multiplicity of each process. }
\end{figure}


As the result analizing 10,000 trajectories, it is found that the
QF process (including a part of DIC), the DQF process and the FF
process account for 72\%, 25\% and 3\%, respectively. In Fig.~3,
the neutron multiplicity from the QF, DQF and FF processes are
denoted by the dashed-dotted line, solid line and gray line,
respectively. The dashed line denotes the total multiplicity
contributed from all processes.

Many trajectories go to the fission area without neutron emission.
The travelling time of such a trajectory is rather short. As can
be seen in Fig.~3, it is difficult to see a characteristic
structure of the neutron multiplicity corresponding to the process
classified. The total number of events for $2n$ emission is
significant, but the QF and the DQF processes account for
approximately equal contributions. These results show that there
is no correlation between the neutron multiplicity and the mass
distribution of the fission fragments in this incident energy.

Figure~4 shows the mass distribution of fission fragments for each
process. The fission fragments from the QF+DIC, DQF and FF
processes are presented by the light gray, the gray and the black shading,
respectively. The thin black line denotes the total number of
events of all processes.
The fission fragments from the FF process spread from $A=100$ to
200. Therefore, we can see that the FF process is related to both
the mass symmetric fission and the mass asymmetric fission. In
Fig.~4, two sharp peaks are located at $A\sim 80$ and $A\sim 210$.
These events come from the trajectories that go down quickly along
the valley at $\alpha \sim 0.42$ in the potential energy surface,
which is the result of the the shell structure of the Pb nucleus.

\begin{figure}
\centerline{
  \includegraphics[height=.37\textheight]{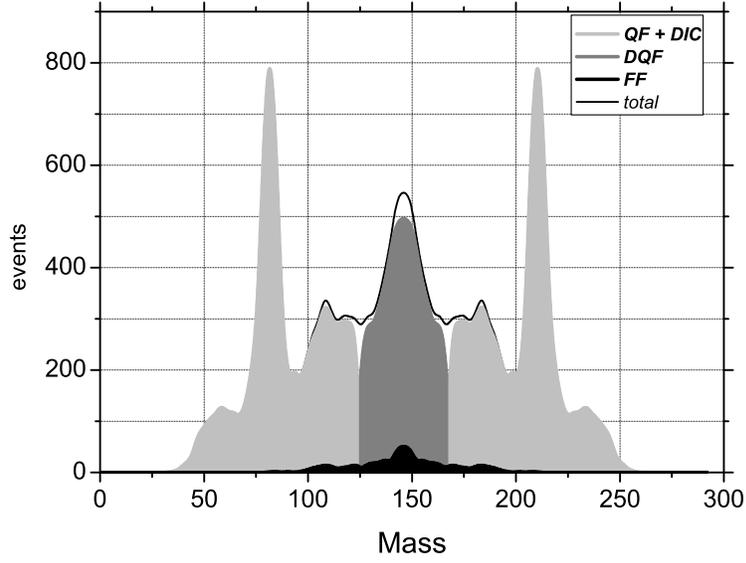}}
  \caption{Mass distribution of fission fragments in the reaction
$^{48}$Ca+$^{244}$Pu at $E^{*}=50$ MeV for $l=0$ using the method
of classification by fission fragment mass. The fission fragments
from the QF+DIC, DQF and FF processes are presented by the light
gray, the gray and the black shading, respectively. The thin black
line denotes the total number of events of all processes.
  }
\end{figure}

\begin{figure}
\centerline{
  \includegraphics[height=.37\textheight]{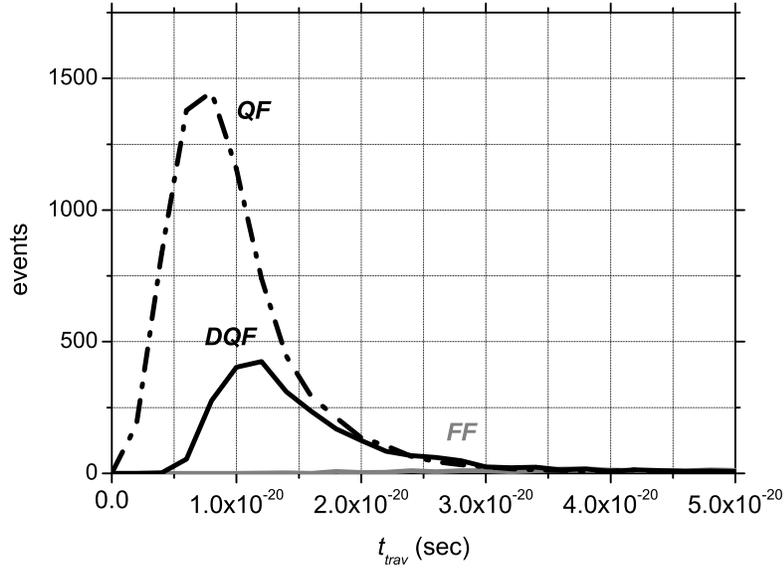}}
  \caption{The distribution of the travelling time $t_{trav}$ of each
  path in the reaction $^{48}$Ca+$^{244}$Pu at $E^{*}=50$ MeV for $l=0$, using the method
of classification by fission fragment mass.}
\end{figure}


The corresponding travelling times for each process are shown in
Fig.~5. The peak of travelling time for the QF process, which is
located at $t_{trav} \sim 7.5 \times 10^{-21}$ sec, is slightly
shorter than the time at the peak of the DQF process. Until
$t_{trav}=3.0 \times 10^{-20}$ sec, the travelling time of DQF
process overlaps with that of the QF process. The average
travelling times of the QF, DQF and FF processes are $1.11, 1.73$
and $4.83 \times 10^{-20}$ sec, respectively. The travelling time
of the FF process is about three times as long as those of the QF
and DQF processes.

\subsubsection{Correlation between the travelling time $t_{trav}$
and the neutron multiplicity $\nu_{n}$}

In the standard classification, we can not distinguish a
characteristic structure of the neutron multiplicity corresponding
to each process, in this low incident energy case. Here, we
investigate the reason why the neutron multiplicity do not show
the characteristic structure corresponding to the reaction process
like as a high incident energy case \cite{ari04a1}. In conclusion,
the cause of this failure is the use of the fission fragment mass
for the classification of processes. We discuss this point in this
subsection.


As we discussed above, we expect that each process may have the
characteristic reaction time. Here, at first we show the relation
between the average travelling time $\langle t_{trav} \rangle$ and
the mass number of the fission fragments in Fig.~6. The average
travelling time of the mass symmetric fission process is longer
than that of the mass asymmetric fission process. We can see
clearly the correlation between the $\langle t_{trav} \rangle$ and
the mass number of the fission fragments. The average travelling
time for $\frac{A_{CN}}{2}-20 < A < \frac{A_{CN}}{2}+20$ is
ranging from 1.8$\times 10^{-20}$ to 2.1$\times 10^{-20}$ sec.
In Fig.~5, the fluctuations of the travelling time of the QF and
the DQF processes, which are expressed by the distributions of the
travelling time around the most probable value for each process,
are ranging in the width of $2.5 \sim 3.0 \times 10^{-20}$ sec.
These fluctuations of the travelling time are larger than the
difference between the average travelling time of the QF and the
DQF processes. Therefore, it is difficult to separate the QF and
DQF processes by the travelling time.

The relation between the travelling time and each neutron
multiplicity in this system is shown in Fig.~7. With increasing
the time duration, the large neutron multiplicity is expected. At
this incident energy, the events for $1n$ and $2n$ emission are
dominant, and it corresponds to the peaks for the QF and DQF
processes in Fig.~3. Due to the stochastic fluctuation in the
statistical model, the time taking for the events of $1n$ and $2n$
emission are overlapped significantly. We can not distinguish the
reaction time of the both processes by the number of neutron
multiplicity.

The life time of the neutron emission and the travelling time are
partly governed by the random number in the stochastic aspect which
is contained in statistical model and the Langevin equation. The
width of these fluctuation in time space is overlapped for each
process in this case. Actually, at $E^{*}=50$ MeV, the maximum
number of evaporated neutron $\nu_{n}^{max}$ is approximately $4\sim
5$ as shown in Fig.~3. However, as shown in Fig.~7, peaks for the
different neutron multiplicity are not well separated so as to
identify the reaction process. As we will show later, it is coming
from small $\nu_{n}^{max}(\sim4)$.

\begin{figure}
\centerline{
  \includegraphics[height=.37\textheight]{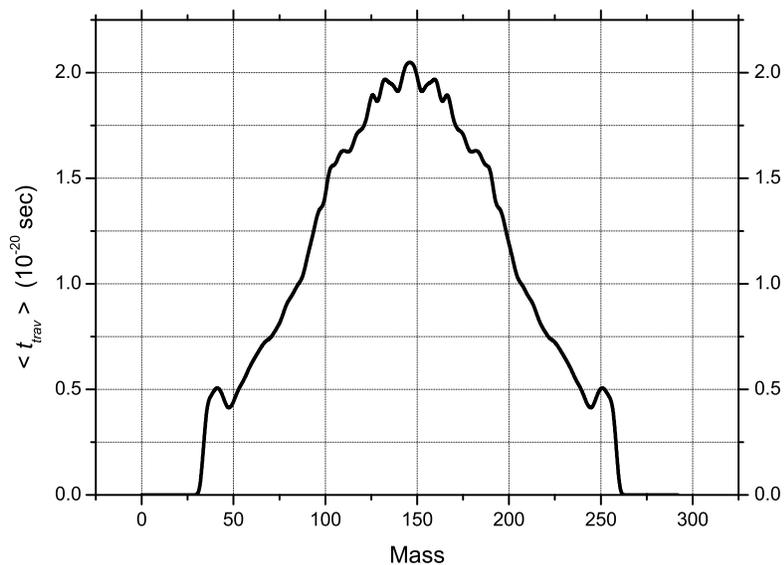}}
  \caption{The average travelling time  associated with fission
  fragment in the reaction $^{48}$Ca+$^{244}$Pu at $E^{*}=50$ MeV for $l=0$.}
\end{figure}

\begin{figure}
\centerline{
  \includegraphics[height=.37\textheight]{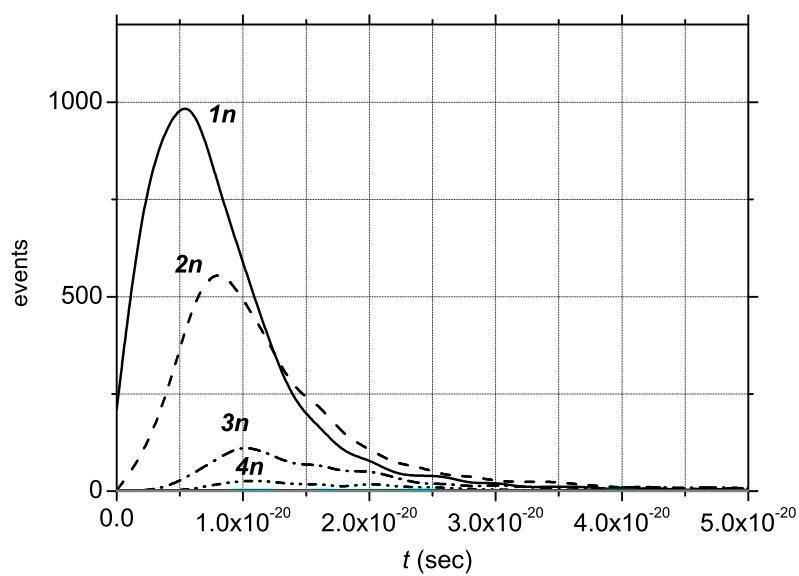}}
  \caption{Time dependence of neutron multiplicity in the reaction
  $^{48}$Ca+$^{244}$Pu at $E^{*}=50$ MeV for $l=0$. The neutron multiplicity are given.}
\end{figure}

\begin{figure}
\centerline{
  \includegraphics[height=.37\textheight]{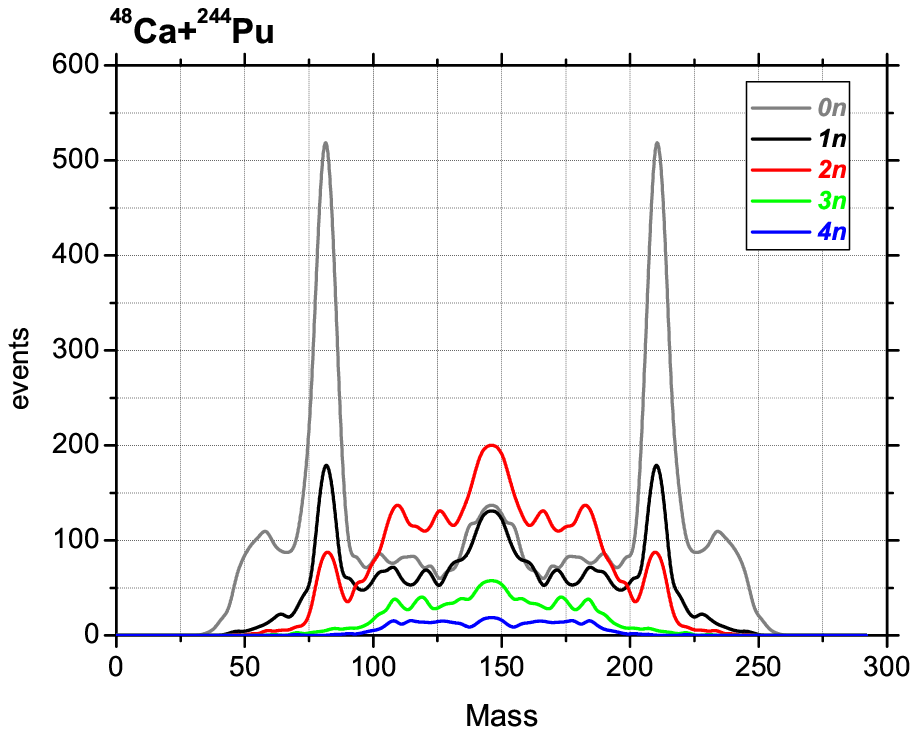}}
  \caption{The neutron multiplicity associated with fission
  fragment in the reaction $^{48}$Ca+$^{244}$Pu at $E^{*}=50$ MeV for $l=0$.}
\end{figure}

Figure~8 shows the neutron multiplicity associated with the mass
number of fission fragments in the reaction $^{48}$Ca+$^{244}$Pu at
$E^{*}=50$ MeV for $l=0$. The two sharp peaks of $0n$ and $1n$ at
$A=80$ and 210 are caused by the trajectory which goes down along
the valley near the Pb nucleus (or the Kr nucleus) in the potential
energy surface. Such a trajectory quickly goes down to the fission
area, so that it does not have large chance to emit any neutron. As
can be seen in Fig. 8, there exist trajectories with a variety of
numbers of the neutron emission in the region $90 < A < 210$. The
trajectories having different neutron multiplicities coexist in this
region. With decreasing mass asymmetry, $2n$ and $3n$ events
increase because, in this region, the number of trajectories with
long travelling time increases as shown in Fig.~6.

The characteristic feature observed in Fig.~8 is that each neutron
multiplicity spreads over a wide mass region of fission fragments.
It is noteworthy that the events with even $0n$ and $1n$ are found
in the mass symmetric fission area. Therefore, it is clear that we
can not extract the characteristic structure of neutron
multiplicity by means of the standard classification.

\subsubsection{Classification using the additional condition on $z$}



As discussed in the previous subsection, the trajectories of each
process spread over a wide mass region of fission fragments. We
understand that each process has characteristic behavior. However,
the information of the mass number of fission fragments does not
clearly connected with the dynamical aspect of each process which
we imaged. We have to look for the proper classification method
that corresponds to our recognition of each process, instead of
the method using fission fragment mass. Here, we propose to employ
a classification method based on the dynamical trajectory of the
process.


As shown in Fig.~3, the QF process includes the events of $4n$
emission. Such events account for about 1.5\% of the total QF
process. Generally, the QF process is regarded as a short process
that does not have enough time to relax the degree of mass
asymmetry. It may be unreasonable to consider that the QF process
has sufficient time to evaporate 4 neutrons.

It is found that among the QF process some trajectories associated
with 4 neutrons emission enter the region of small $z$ and their
travelling times seem longer than that of the typical sample
trajectory presented in our previous paper \cite{ari04a}. These
trajectories pass the scission points where the mass number is
greater than $\frac{A_{CN}}{2}+20$ $ (\alpha=0.136)$ and less than
$\frac{A_{CN}}{2}-20$ $ (\alpha=-0.136)$, which is classified into
the QF process by the standard classification. In particular, it
is found that the mass asymmetry of such trajectory changes from
$\alpha =0.67$ to $-0.3$ in the region of small $z$. This QF
process should be distinguished from the usual QF process.

We usually consider the trajectory of the DQF process to enter the
small $z$ region, but due to the strong Coulomb repulsion, it goes
to the large-deformation region and returns to the fission area
without passing the fusion box. Thus, considering this scenario,
we add a condition to the standard classification discussed in the
previous subsection.

Therefore, we define a new rule that a trajectory which enters the
region of $z < 0.4$, where the nuclear shape resembles
mono-nucleus, is to be reclassified into the DQF process, even
though the mass fragments lie in the asymmetric region ($A <
\frac{A_{CN}}{2}-20$ and $A > \frac{A_{CN}}{2}+20$). That is to
say, the DQF process is redefined as the process that the
trajectory enters the region of $z < 0.4$ without passing the
fusion box. Actually, the distribution of the travelling time of
such a trajectory occupies the region of longer $t_{trav}$ of the
QF process. We call this classification "the classification of the
fission fragment mass with an additional condition on  $z$".


Here, we should consider the process that the trajectory goes to
mass symmetric fission region, where the mass number is greater
than $\frac{A_{CN}}{2}-20$ and less than $\frac{A_{CN}}{2}+20$,
but the trajectory does not enter the region of $z < 0.4$. We name
this process the mass symmetric quasi-fission process (MSQF). In
this system, the trajectory of the MSQF process occupies 1.21 \%
of the all trajectories, and $4.55 \%$ of the mass symmetric
fission events. We can say that the almost mass symmetric fission
events come from the trajectory which enters the region of $z <
0.4$.

\begin{figure}[h]
\centerline{
  \includegraphics[height=0.39\textheight]{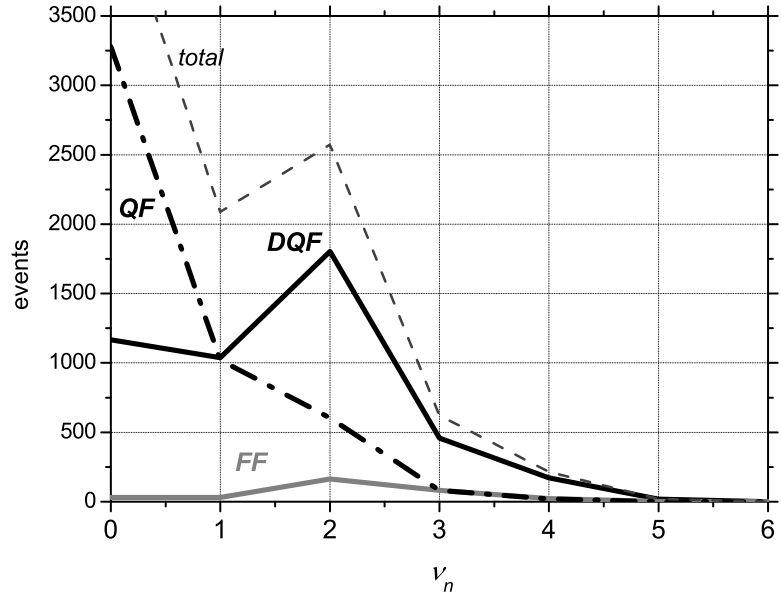}}
  \caption{Pre-scission neutron multiplicity given by our theoretical calculation
   in the reaction $^{48}$Ca+$^{244}$Pu at $E^{*}$=50 MeV for $l=0$ using the method of
   classification by fission fragment mass adding a condition on $z$.
   The lines denote the same items as those in Fig.~3. }
\end{figure}

\begin{figure}
\centerline{
  \includegraphics[height=.37\textheight]{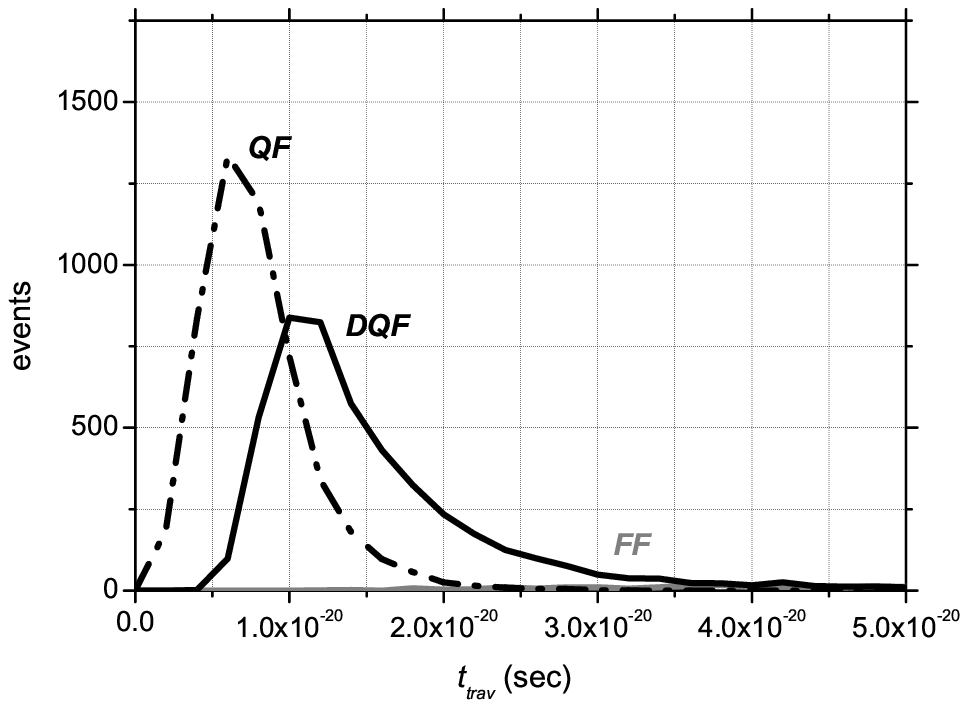}}
  \caption{The distribution of the travelling time $t_{trav}$ of each path in
   the reaction $^{48}$Ca+$^{244}$Pu at $E^{*}=50$ MeV for $l=0$, using the method
   of classification by fission fragment mass adding a condition on $z$.}
\end{figure}

\begin{figure}
\centerline{
  \includegraphics[height=.37\textheight]{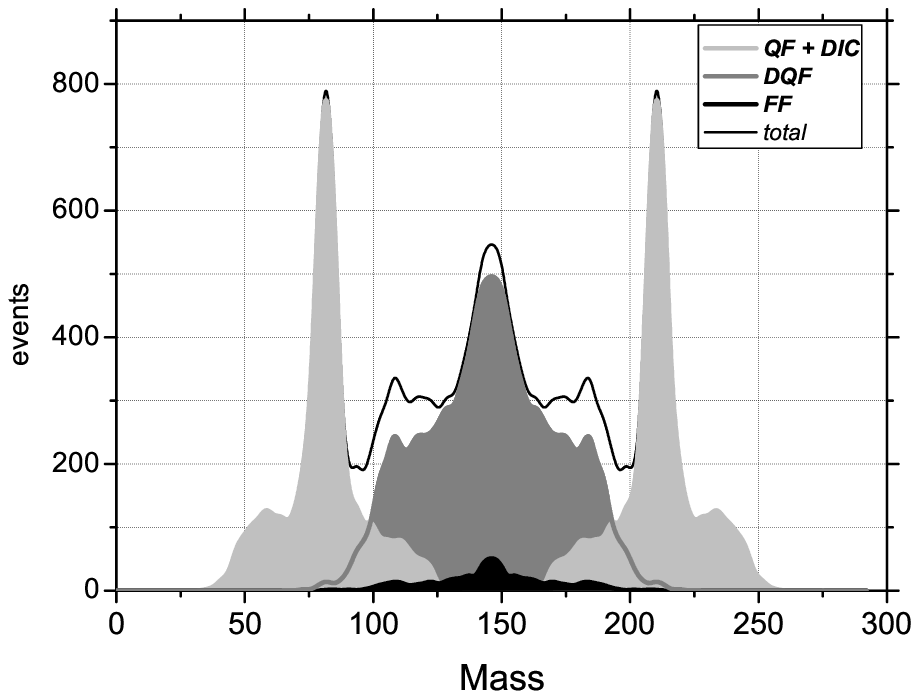}}
  \caption{Mass distribution of fission fragments in the reaction
$^{48}$Ca+$^{244}$Pu at $E^{*}=50$ MeV for $l=0$ using the method
of classification by fission fragment mass adding a condition on
$z$. The shading and lines denote the same items as those in
Fig.~4. }
\end{figure}


By this classification, the component of the last QF process,
which has a long travelling time, is classified into the DQF
process. Such a trajectory accounts for about 32\% of the total QF
process. Figure~9 shows the each neutron multiplicity distribution
by the new classification, where each line has the same meaning as
that in Fig.~3. We can see that the peak of the DQF process at
$\nu_{n}=2$ is higher and the number of events increases compared
with that in Fig.~3.
Figure~10 shows the distribution of
$t_{trav}$ for each process using this new classification. The the
travelling time of the DQF process and the QF process separates
more clearly than that shown in Fig.~5. After $t_{trav}=1.2 \times
10^{-20}$ sec, the DQF process is dominant.

The mass distribution of fission fragments is shown in Fig.~11.
The shading and lines have the same meanings as those in Fig.~4.
We can see that the fission fragments from the DQF process exist
in the region of mass number greater than $\frac{A_{CN}}{2}+20$ $
(A=166)$ and less than $\frac{A_{CN}}{2}-20$ $ (A=126)$. In the
regions $166< A < 200$ and $90 <A <126$, the DQF process and the
QF process coexist. This means that the trajectories with long and
short travelling times intermingle. That is to say, in this
region, the fission fragment has two possible origins, one is the
short process wherein the colliding partner goes  quickly to the
fission area, and the other is the long process in which the
trajectory approaches the compact shape and takes a long time to
arrive at the fission area.


\subsubsection{Incident energy dependence in neutron multiplicity
analysis}


As practice, in order to identify the details of the whole
reaction process, we introduce the classification on the basis of
trajectory's behavior by the theoretical calculation. When the
experimental pre-scission neutron multiplicity is analyzed in
combination with the trajectory calculation, the information on
the neutron multiplicity can be put to practical use to clarify
the reaction mechanism, that is to say we can distinguish the QF,
DQF and FF processes.

At a high incident energy, the situation is different. When the
incident energy increases, the distribution of the neutron
multiplicity is spread well. The neutron multiplicity can become a
good measure of each process
\cite{ari04a1,ari04a2,ari04a3,ari04a0}.

Figure~12 shows the distribution of the neutron multiplicity in
the reaction $^{48}$Ca+$^{244}$Pu at $E^{*}= 80$ MeV and $E^{*}=
160$ MeV for $l=0$ in (a) and (b), respectively. Here, each line
has the same meaning as that in Fig.~3. We use the classification
of the fission fragment mass with an additional condition on $z$.
With increasing the incident energy, the events of the QF process
decrease because the mass symmetric fission events are dominant.
The DQF and FF processes are well distinguished by the neutron
multiplicity at $E^{*}=160$ MeV.

In the high incident energy case, $\nu_{n}^{max}$ increases, which
are about 8 and 14 at $E^{*}=80$ and 160 MeV, respectively. With
increasing $\nu_{n}^{max}$, it becomes possible to identify the
reaction processes using the distribution of neutron multiplicity,
because the neutron multiplicity spreads over the wide region and
indicates the characteristic property of each process, which is
not overlapped by the fluctuations.

\begin{figure}[h]
\centerline{
  \includegraphics[height=0.65\textheight]{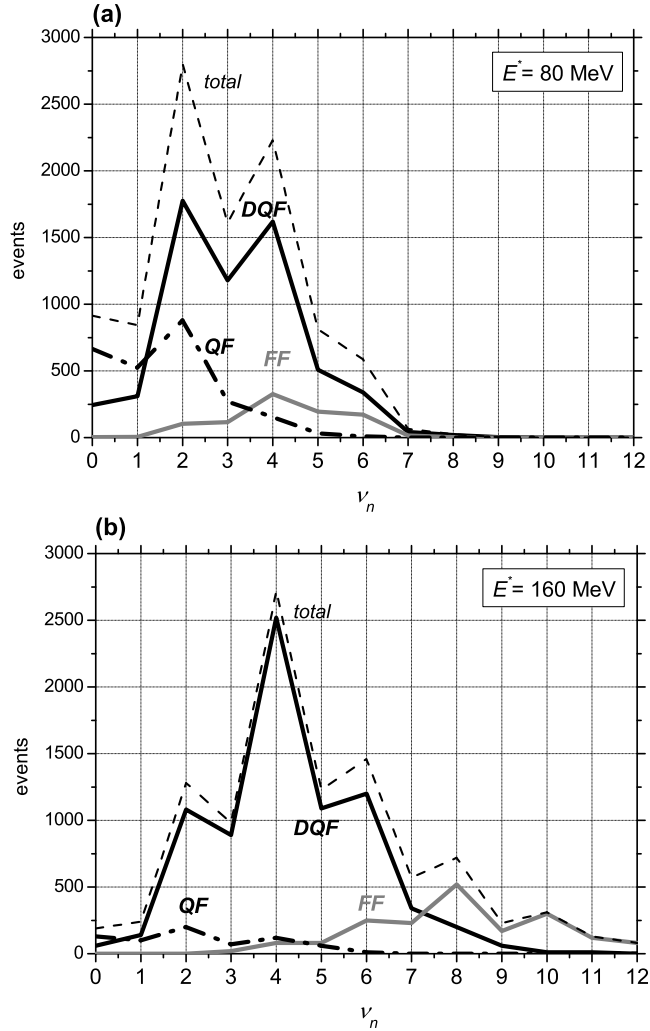}}
  \caption{Pre-scission neutron multiplicity given by our theoretical calculation
   in the reaction $^{48}$Ca+$^{244}$Pu at (a)~$E^{*}$=80 MeV and (b)~~$E^{*}$=160 MeV for $l=0$,
    using the method of classification by fission fragment mass adding a condition on $z$.
   The lines denote the same items as those in Fig.~3}
\end{figure}

Table~1 shows the average travelling time $\langle t_{trav}
\rangle$ in the reaction $^{48}$Ca+$^{244}$Pu for $l=0$, using the
classification method (i)~the fission fragment mass (method~(i))
and (ii)~the fission fragment mass with additional condition on
$z$ (method~(ii)). At $E^{*}=50$ MeV, in method~(i), the $\langle
t_{trav} \rangle$ of the DQF process is about 1.5 times as long as
that of the QF process. However, in method~(ii), it is about twice
as long. This is because the DQF process in method~(ii) includes
the long QF process in method~(i). The $\langle t_{trav} \rangle$
of the DQF process in method~(ii) does not change as much as that
in method~(i), because the long tail of the DQF process mainly
influences the average time.

The $\langle t_{trav} \rangle$ at $E^{*}=80$ MeV and 160 MeV are
shown in Table~1, in the classification method (ii). We can see
the incident energy dependence of $\langle t_{trav} \rangle$. With
increasing the incident energy, the initial speed of the
trajectory increases. After the kinetic energy dissipates, also
the moving speed of trajectory by the fluctuation is high. As a
result, $\langle t_{trav} \rangle$ becomes shorter. The ratio
among the $\langle t_{trav} \rangle$ of QF, DQF and FF seems to be
approximately constant for each incident energy.


\vspace{1.5cm}
\begin{table}[h]
\begin{center}
\caption{ The average travelling time for each classification, in
the reaction $^{48}$Ca+$^{244}$Pu for $l=0$. The methods of
classification are based on (i)~the fission fragment mass
(method~(i)) and (ii)~the fission fragment mass with a condition
on $z$ (method~(ii)).} \vspace{0.5cm}

\begin{tabular} {|c|r|r|r|r|} \hline\hline
  &\multicolumn{4}{c|}{
$^{48}$Ca+$^{244}$Pu} \\ \hline
 &method~(i)&\multicolumn{3}{c|}{method~(ii)}\\ \hline
 &$E^{*}=50$ MeV&$E^{*}=50$ MeV&$E^{*}=80$ MeV&$E^{*}=160$ MeV \\ \hline
 & ($\times 10^{-20}$sec)& ($\times
 10^{-20}$sec)& ($\times 10^{-20}$sec) &($\times 10^{-20}$sec) \\
 {\em QF} &1.11&0.88&0.85&0.73 \\
 {\em DQF} &1.73&1.69&1.58&1.38 \\
 {\em FF}  & 4.83 & 4.83 &4.00&3.71\\ \hline\hline

\end{tabular}
\end{center}
\end{table}

\vspace{1.5cm}

\section{Summary}

The fusion-fission process in the superheavy-mass region was
studied on the basis of fluctuation-dissipation dynamics.
In order to classify the reaction process in greater detail, we
analyzed the pre-scission neutron multiplicity connecting with the
mass distribution of fission fragments. The neutron multiplicity
depends on the travelling time of the trajectory from the point of
contact to the scission point. We introduced the effect of
pre-scission neutron emission into the three-dimensional Langevin
calculation, that is to say, we combined the Langevin code with
the statistical code. We applied our model to the investigation of
the whole reaction process in the reactions $^{48}$Ca+$^{208}$Pb
and $^{48}$Ca+$^{244}$Pu at the incident energy corresponding to
the excitation energy of a compound nucleus, $E^{*}= 50$ MeV.

In the reaction $^{48}$Ca+$^{208}$Pb at $E^{*}= 50$ MeV, the
reaction process is simple. Most of the process comprises the FF
process. In this case, the pre-scission neutron multiplicity
distribution shows a single peak at $\nu_{n}=2$.

In the reaction $^{48}$Ca+$^{244}$Pu at $E^{*}= 50$ MeV, the
pre-scission neutron multiplicity distribution of the DQF process
overlaps significantly with the distribution of the QF process,
when we use the standard classification using only the information
of mass distribution of fission fragments. Consequently, we could
not distinguish between the QF and the DQF processes on the basis
of the pre-scission neutron multiplicity. However, if we introduce
the additional condition for the classification concerning with
the travelling time of trajectories, we could show that the
situation becomes to be resolved. The additional condition is that
the trajectory entering into the area of $z < 0.4$ but not into
the fusion box is classified to DQF process, even if the mass
fragments distribute outside the region between
$\frac{A_{CN}}{2}-20$ and $\frac{A_{CN}}{2}+20$. Using this
classification, it was shown that the pre-scission neutron
multiplicity distribution of the DQF process is clearly separated
from that of the QF process.

We also discussed how the characteristic structure of the neutron
multiplicity depends on the incident energy, and presented that we
can get more and more clear structure corresponding to each
reaction process when the incident energy increases.

In the next study, we would like to investigate parameters more
precisely, for example, the friction tensor, the level density
parameter and the neutron binding energy near di-nucleus
configuration etc. We try to use the friction tensor which is
derived from liner response theory at a low excitation energy
\cite{hofm97,yama97,ivan97}.



\section*{Acknowledgement}

The authors are grateful to Professor Yu.~Ts.~Oganessian, Professor
M.G.~Itkis, Professor V.I.~Zagrebaev for their helpful suggestions
and valuable discussion throughout the present work. The authors
wish to thank Professor T.~Wada who developed the original version
of the calculation code for the three-dimensional Langevin equation.
The special thanks are deserved to Dr. A.K.~Nasirov for useful
discussion. The authors thank Dr. S.~Yamaji and his collaborators,
who developed the calculation code for potential energy with a
two-center parameterization. This work has been in part supported by
INTAS projects 03-01-6417.

\end{document}